# Differential Evolution Approach to Detect Recent Admixture

Konstantin Kozlov, Dmitry Chebotarov, Mehedi Hassan, Martin Triska, Petr Triska, Pavel Flegontov, Tatiana Tatarinova

Feb 20, 2015

## Abstract

The genetic structure of human populations is extraordinarily complex and of fundamental importance to studies of anthropology, evolution, and medicine. As increasingly many individuals are of mixed origin, there is an unmet need for tools that can infer multiple origins. Misclassification of such individuals can lead to incorrect and costly misinterpretations of genomic data, primarily in disease studies and drug trials. We present an advanced tool to infer ancestry that can identify the biogeographic origins of highly mixed individuals. *reAdmix* can incorporate individual's knowledge of ancestors (e.g. having some ancestors from Turkey or a Scottish grandmother). *reAdmix* is an online tool available at http://chcb.saban-chla.usc.edu/reAdmix/.

## Background

The ability to trace individuals to the point where their DNA was formed at the population level poses a formidable challenge in genetic anthropology, population genetics and personalized medicine [1]. The vast progress accomplished in developing resources for identifying candidate gene loci for medical care and drug development [2] was largely unmatched by the field of biogeography and ancestral inference. Only in the past decade have researchers begun harnessing high-throughput genetic data to improve our understanding of global patterns of genetic variation and its correlation to geography. This is not surprising, because the genetic variation is largely determined by demographic history of inbreeding or admixture which often vary between geographic regions. Although in the past few years we have witnessed a growing interest in biogeography methods, only a few computational tools exist, particularly for analysis of mixed individuals [3, 4, 5, 6].

These methods can be either local (focusing on origin of chromosomal segments), such as Lanc-CSV [7], LAMP-LD [8], and MULTIMIX [9], global (average ancestral proportions across the genome), such as ADMIXTURE [10], STRUCTURE [11, 12], or both, such as HAPMIX [13], LAMP [8, 14]. Some popular applications are PCA-based [3]. For humans, PCA was shown to be accurate within 700 kilometers in Europe [3]. The Spatial Ancestry Analysis (SPA) [4] is an advanced tool that explicitly models allele frequencies. However, estimated by the percentage of individuals correctly assigned to their country of origin, the accuracy of both PCA and SPA remain low for Europeans (40 ± 5% and 45 ± 5%, respectively) and are even less for non-Europeans [4], suggesting their limitation for biogeographic applications [4, 15, 16]. Note, that the country of origin does not necessarily correlate with ethnicity. SPAMIX [17] is reported to have an accuracy of 550Km for two-ancestral admixtures, which is impressive but insufficient. Algorithms like mSpectrum [18], HAPMIX [13] and LAMP [8] achieve good accuracy at a continent resolution [18], but do not achieve country-level resolution. Related tools like BEAST [19], STRUCTURE [12], and Lagrange [20] are either inapplicable to autosomal data or cannot be used to study recent admixture in humans, animals, and plants. We note that looking at Y chromosome and mtDNA alone is insufficient for detailed biogeographic analysis, since closely related populations have similar distributions of haplogroups.

To address these limitations, we have recently developed an admixture-based tool, Geographic Population Structure (GPS), that can accurately infer ancestral origin on unmixed individuals [21]. GPS infers the geographical origin of individual by comparing the his/her "genetic signature" to those of reference populations known to exhibit low mobility in the recent past. GPS's accuracy was demonstrated by classifying 83% worldwide individuals to their country of origin and 65% to a particular region of the country. Applied to over 200 Sardinian villagers, GPS placed 25% of them in their villages and ≈ 50% within 50 kilometers of their villages.

However, contemporary individuals often migrate to different areas and bear offspring of mixed geographical origins. GPS would incorrectly predict such offspring to the central point between the parental origins, which would be unsuitable for pharmacology, forensics, and genealogy; therefore, GPS is not equipped to handle mixed individuals. Moreover, often individuals have an indication of at least one of their possible origins, which can be used to improve the prediction, but existing tools are not designed to consider such information. To address these limitations, we propose *reAdmix*, a novel tool that models individuals as a mix of populations and can use user input to improve its predictions. We demonstrate the accuracy of *reAdmix* on a simulated dataset and compare its performance with three alternative tools. *reAdmix* can be useful for professionals trying to match cases and controls in disease studies, scientists studying bio-diversity and origins of humans, animals, and plants, as well as many people seeking answers about their past.

# Results and discussion

*reAdmix* expands the admixture based approach, described in [21]. It requires building a dataset of worldwide populations (*reference set*), by applying an unsupervised ADMIXTURE [10] analysis with various number of components. As shown in Elhaik et al. [21], the most suitable number of components was verified using a PCA-based analysis. After choosing an optimal number of ancestral populations, $K$, allele frequencies inferred for each of the ancestral populations with ADMIXTURE formed a reference dataset for subsequent steps. Individuals were projected onto this reference dataset of $K$ ancestral populations using ADMIXTURE in a supervised mode. In other words, an individual's genotype was "broken down" into a predefined set of ancestral components. These admixture proportions represent a tested individual in the space of $K$ putative ancestral populations (for example, in case of $K = 9$, the ancestral populations are North-East Asian, Mediterranean, South African, South-West Asian, Native American, Oceanian, South-East Asian, Northern European, Sub-Saharan African). Details of the admixture components calculations are described in the Methods section. The task of *reAdmix* is to present individual's ancestry as a weighted sum of modern reference populations (e.g. 25% French, 25% German, 50% Japanese) based on these $K$ admixture components. The goal is to find the smallest number of reference populations that represent the tested individual with the highest possible accuracy. We used the reference population panel with known admixture components relative to putative-ancestral population. Preparation of this dataset is described in the Methods section of this manuscript.

*reAdmix* can operate in unconditional (nothing is known about the tested individual) and conditional (there is partial information about individual's ancestors) modes. If the prior information contradicts the individual's genotype, it is discarded. See Methods for detailed description of the *reAdmix* approach.

Briefly, the tested individual and the $N$ reference populations are represented as points inside the standard simplex in $K$-dimensional space, via their $K$ admixture coefficients. For example, the genome of an individual that consists of 50% population $X$, 25% population $Y$, and 25% population $Z$ can be represented by the corresponding point $T$ as a convex combination:

$$T = 0.5X + 0.25Y + 0.25Z,$$

where each population is represented by a vector of $K$ admixture coefficients, for example:

$$X = [0.1, 0.15, 0.25, 0, 0, 0.5, 0, 0, 0].$$

Thus, the question of determining the population mixture of an individual, i.e. the parental populations and their proportions, can be translated into the following problem in the $K$-dimensional admixture space: find a representation of a given test point as a convex combination of a subset of $N$ reference points.

Note that both test and reference points have the property that their coordinates, being admixture proportions, sum to one; therefore they belong to the standard $(K - 1)$-dimensional simplex defined by the equation $\sum_{k=1}^{K} x_k = 1$. The set of all convex combinations of the $N$ reference population points (their *convex hull*) is a polytope, a higher-dimensional analogue of polyhedron, inside the standard simplex. Our problem has a solution if the test point is located inside this polytope. The solution is not necessarily unique: when $N$ exceeds $K + 1$, the point can be represented by several convex combinations of reference populations. Hence, there are multiple mixture combinations can explain the individual's admixture. One way to get parametric uniqueness is to find the smallest dimension simplex containing the given point and reduces the combinatorial freedom. Although there may still be many simplices of the same dimension containing the same point, it becomes unlikely when the dimension of the ambient space gets higher. Another way is to take advantage of prior information provided by the user (e.g. if the individual knows some of his/her ancestry).

We conducted several tests of *reAdmix* accuracy described below. The tests were performed on the computer with Intel Xeon 2x5650@2.67GHz CPU (24 cores HT), 24 Gb RAM, and took about 50 sec and 40 Mb RAM per one sample. In optimization runs, five worker threads were employed in parallel.

## Comparison with GPS using unmixed individuals

To test the performance of *reAdmix* we first applied it to worldwide unmixed samples, whose admixture coefficients were averaged over individuals with the same self-reported origin. The program was tested under two conditions: either no prior information or random incorrect prior information was supplied. *reAdmix* correctly identified the individuals as unmixed in 96% and 86% for these experiments, respectively. Two scores were then computed: percent of individuals matching the correct population and distance to correct population. *reAdmix* correctly determined the population of 96% of the samples. The incorrectly predicted individuals were placed within an average distance of 35 kilometers to their reported location. When incorrect prior information was provided, the quality did not drop drastically: 88% of samples was mapped to the reported population, with an average distance of 165 kilometers to the correct geographical location. These results indicate the robustness of *reAdmix*.

## Simulated marriages

Next, we simulated multiple mixture scenarios and tested the ability of *reAdmix* to correctly identify the populations in each mixture and their mixture proportions. We considered several relevant scenarios for an American of a European descent where individuals may have two, three or four European/Near Eastern origins and tested the ability of *reAdmix* to correctly identify the populations and proportions in simulated mixed families. These mixtures are currently common for big cities in North America. Individuals of mixed origin were simulated from admixture vectors of un-mixed individuals. For each of the three scenarios, we randomly generated 300 family structures by sampling from population means from different populations in the reference dataset and computed the weighted average of their corresponding admixture coefficients with varying error term:

$$T = \sum w_i \times r_i + \epsilon \times N \left( mean = 0, st.dev = \sum w_i \times \sigma(r_i) \right),$$

where $\epsilon$ is the scaling parameter and the error is normally distributed with zero mean and the standard deviation equal to the weighted sum of deviations for mixture components. Notice, that admixture vectors do not contain chromosomal positions, and, therefore, information about haplotype blocks is not utilized in our approach.

We tested the algorithm in unconditional and conditional modes. A single correct population was provided for the tests of the conditional mode. We also tested the case in which the mixture weights are known to be equal *a priori*. Our simulation results are shown in Tables 1-2. The scenarios are named according to the percentage of mixed ancestral population, e.g. "50x50". The "Correct position" is defined as a prediction within 320 km of the reported location. The number of cases with at least one correctly predicted origin in conditional mode gives the number of cases in which the unknown population is also predicted correctly, and hence it can be less than the number of correctly predicted positions. Conditioning on one population reduces the average distance to correct population more than two-fold.

Next, in order to represent an increasing trend of marriages between spouses of a different ethnicity we added several Native American populations. The most common type of cross-ethnic marriages in the US is European/Latino couples, accounting for 43% of cross-ethnic marriages [22]. Due to the sparse coverage of Amerindians and the large geographic distances between populations compared to European ones, we expected a significant decline in *reAdmix* performance, however, the decline was less severe than expected (Tables 3-4).

## Testing the four-way admixtures

Finally, we compared *reAdmix* to mSpectrum[18], HAPMIX [13] and LAMP [8, 14] programs. We used the benchmark of Sohn et al.[18]. In this benchmark, four-way admixtures were generated using Russian, Bantu Kenya, Pima, and Yi populations in proportions $\eta(1) = (0.2, 0.8, 0, 0)$ and $\eta(2) = (0.8, 0.15, 0.03, 0.02)$. This corresponds to (19.8 : 80.2 : 0 : 0) and (83.3 : 13.1 : 1.5 : 2.1) in the space of European, African, Native American and East Asian ancestries. Tables 6 and 5 and Figure 1 show comparative performance of the four methods using the two- and four-way admixed individuals. Proportions determined by *reAdmix* (in unconditional mode) were the closest to the true mix of ancestries. In case of two-ways admixed individuals *reAdmix* in unconditional mode was able to determine not only the continent of origin, but the precise population mix (Russian and Bantu Kenya) and proportions (0.2 and 0.8). In case of the four-ways admixed individuals, there are 2317 different ethnic composition at a country level with the same admixture composition in the space of European, African, Native American and East Asian ancestries. Therefore, selection of the "best" ethnic composition is intrinsically difficult or even impossible when the number of components (K) is small and the mixture is complex. In our web application we use larger values of *K*.

Table 1: Accuracy of *reAdmix* ancestry predictions for different mixture scenarios from European populations. Percentage of mixed ancestral population is given in the "Scenario" column. "Correct position" is defined as a prediction within 320 km of reported location. "Correct populations" is defined as a geographically correct prediction where the method correctly discriminated between neighboring populations.

| Scenario | Prior | Correct position (%) | At least one correctly predicted origin (%) | Correct populations (%) | Average distance to correct population, km |
|---|---|---|---|---|---|
| 50x50 | none | 100 | 83 | 16 | 505 |
| | 1 pop. | 100 | 75 | 31 | 8 |
| | equal weight | 100 | 81 | 26 | 251 |
| 50x25x25 | none | 98 | 80 | 1 | 572 |
| | 1 pop. | 100 | 61 | 2 | 240 |
| 25x25x25x25 | none | 99 | 79 | 0 | 729 |
| | 1 pop. | 100 | 61 | 0 | 427 |

## Applicability to other species

*reAdmix* can be applied to analyze geographic origin of other species, provided there is a sufficient collection of ancestry-informative markers for the organism of interest. Elhaik et al [21] estimated that thinning of the 150,000 Geno2.0 set of markers to 40,000 randomly selected SNPs resulted in 3% error in admixture coefficients. In order to justify usage of even smaller genotyping datasets, we calculated the expected bias from supplementing the reference set with admixture components of populations genotyped over fewer markers down to randomly selected 500 markers. For that, we randomly selected 500 markers for nine populations from 1000 genomes dataset, and generated admixture proportions using ADMIXTURE program. The resulting proportions were compared to those obtained using the complete marker set. We found very small differences in the admixture proportions that slowly increased for thinner marker sets. Even with 500 markers, the largest observed difference (6%) was within the within-variation of our populations and did not affect the assignment accuracy. These results confirm the robustness of admixture-based approach and its usability for datasets as small as 500 ancestry informative markers (markers whose frequencies are significantly different, between two or more populations). We are currently developing *reAdmix* portals for *Arabidopsis thaliana*, *Medicago truncatula*, *Oryza sativa*, *Elaeis guineensis* and *Drosophila melanogaster*.

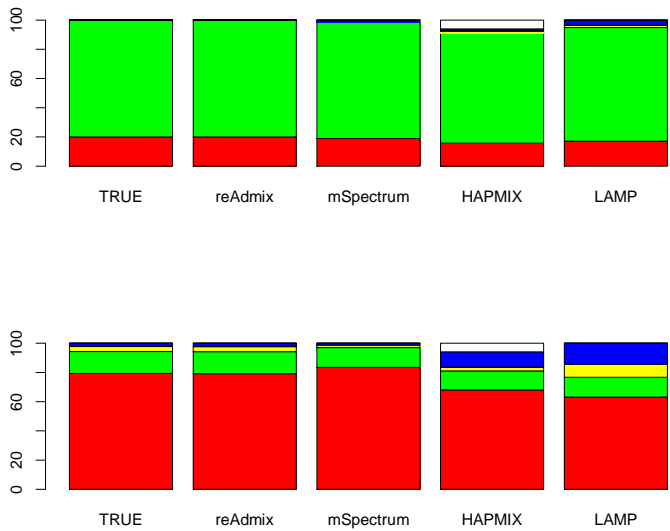

Figure 1: Performance of *reAdmix*, mSpectrum, HAPMIX and LAMP using two-way (top) and four-way (bottom) admixed individuals. Color coding: red - European, green - African, yellow - Native America, blue - East Asian, and white - unassigned.

Table 2: Accuracy of *reAdmix* ancestry predictions for different mixture scenarios from European populations with error term, $\epsilon$, to simulate variability of admixture proportions within populations. Percentage of mixed ancestral population is given in the "Scenario" column. "Correct position" is defined as a prediction within 320 km of reported location. "Correct populations" is defined as a geographically correct prediction where the method correctly discriminated between neighboring populations.

| Scenario | Error, $\epsilon$ | Correct position (%) | At least one correctly predicted origin (%) | Correct populations (%) | Average distance to correct population, km |
|---|---|---|---|---|---|
| 50x50 | 0.01 | 99 | 72 | 6 | 401 |
|  | 0.03 | 99 | 74 | 5 | 363 |
|  | 0.05 | 99 | 73 | 5 | 386 |
| 50x25x25 | 0.01 | 99 | 81 | 0 | 588 |
|  | 0.03 | 99 | 79 | 0 | 553 |
|  | 0.05 | 98 | 79 | 0 | 557 |
| 25x25x25x25 | 0.01 | 99 | 81 | 0 | 600 |
|  | 0.03 | 98 | 78 | 0 | 618 |
|  | 0.05 | 98 | 80 | 0 | 623 |

Table 3: Accuracy of *reAdmix* ancestry reconstruction for different mixture scenarios from European and Native American populations. Percentage of mixed ancestral population is given in the "Scenario" column. "Correct position" is defined as a prediction within 320 km of reported location. "Correct populations" is defined as a geographically correct prediction where the method correctly discriminated between neighboring populations.

| Scenario | Condition $i$ | Correct position (%) | At least one correctly predicted origin (%) | Correct populations (%) | Average distance to correct population, km |
|---|---|---|---|---|---|
| 50x50 | none | 98 | 89 | 30 | 329 |
|  | 1 pop. | 99 | 87 | 36 | 2 |
|  | equal weight | 99 | 88 | 36 | 135 |
| 50x25x25 | none | 86 | 81 | 18 | 1390 |
|  | 1 pop. | 94 | 72 | 4 | 362 |
| 25x25x25x25 | none | 86 | 85 | 0 | 1484 |
|  | 1 pop. | 90 | 71 | 0 | 759 |

Earlier [21], we demonstrated that sample sizes used to generate database reference populations varied between $N = 2$ and $N = 15$ and were not correlated with prediction accuracy ($r = 0.01$). For well covered areas, the sizes can be as small as $N = 2$. Note, that a fully sequenced genome is not required for *reAdmix* method, only a collection of SNPs. This extends the applicability of the *reAdmix* to species with limited genomic information.

Table 4: Accuracy of *reAdmix* ancestry predictions for different mixture scenarios from European and Native American populations with error term, $\epsilon$, to simulate variability of admixture proportions within populations. Percentage of mixed ancestral population is given in the "Scenario" column. "Correct position" is defined as a prediction within 320 km of reported location. "Correct populations" is defined as a geographically correct prediction where the method correctly discriminated between neighboring populations.

| Scenario | Error, $\epsilon$ | Correct positions (%) | At least one correctly predicted origin (%) | Correct populations (%) | Average distance to correct population, km |
|---|---|---|---|---|---|
| 50x50 | 0.01 | 97 | 83 | 12 | 354 |
| | 0.03 | 97 | 83 | 9 | 391 |
| | 0.05 | 98 | 84 | 7 | 357 |
| 50x25x25 | 0.01 | 88 | 80 | 2 | 1156 |
| | 0.03 | 85 | 77 | 2 | 1254 |
| | 0.05 | 88 | 81 | 1 | 1147 |
| 25x25x25x 250.03 | 0.01 | 85 | 82 | 0 | 1554 |
| | | 85 | 82 | 0 | 1526 |
| | 0.05 | 87 | 82 | 0 | 1441 |

# Conclusions

The ability to identify the geographic origin of an individual using genomic data poses a formidable challenge due to its complexity and potentially dangerous misinterpretations [23]. Knowledge of biogeography and recent ancestry are essential for research in multiple fields such as biodiversity, genealogy, anthropology, sociology, and forensics, as well as personalized medicine and epidemiology in which ancestry is an important covariate. Development of *reAdmix* is a response to the high demand for improved and accurate ancestry identification methods, it can accurately measure admixture and infer biogeography in complete-genome data sets that are now practical to generate. *reAdmix* is a computationally efficient and organism-independent tool that can be easily applied to a variety of species where sufficient collection of ancestry-informative markers are available. We expect to improve perfor-mance of *reAdmix* with inclusion of additional world-wide reference samples and further computational development.

| Ethnicity | True | ReAdmix | mSpectrum | HAPMIX | LAMP |
|---|---|---|---|---|---|
| European | 79.3 | 79.2 | 83.5 | 68.1 | 63.2 |
| African | 15 | 15 | 13.5 | 13 | 13.5 |
| Nat. American | 3.5 | 3.5 | 2.6 | 2.6 | 8.9 |
| East Asian | 2.2 | 2.3 | 0.4 | 10.4 | 14.4 |
| Other | 0 | 0 | 0 | 5.9 | 0 |

Table 5: Performance of *reAdmix*, mSpectrum, HAPMIX and LAMP using four-way admixed individuals. Estimation errors for the four-way admixture were 0.10, 4.89, 15.24, 20.96, respectively.

| Ethnicity | True | ReAdmix | mSpectrum | HAPMIX | LAMP |
|---|---|---|---|---|---|
| European | 20 | 20 | 18.9 | 15.7 | 17.1 |
| African | 80 | 80 | 79.5 | 76.7 | 77.8 |
| Nat. American | 0 | 0 | 1.2 | 0.3 | 1.6 |
| East Asian | 0 | 0 | 0.4 | 1.3 | 3.5 |
| Other | 0 | 0 | 0 | 6 | 0 |

Table 6: Performance of *reAdmix*, mSpectrum, HAPMIX and LAMP using two-way admixed individuals. Estimation errors for the two-way admixture were 0.01, 1.70, 8.18, and 5.28, respectively.

## Methods

### Reference database

150K Dataset

To create a reference set we used 600 worldwide individuals collected as part of the *Genographic Project* and the *1000 Genomes Project* and genotyped on the GenoChip [24], containing 150K ancestry-informative markers, and 1043 *Human Genome Diversity Project* (HGDP) samples genotyped on Illumina 650Y array, containing 661K markers. SNP marker set of the GenoChip array (Genographic Project) was selected as a basic one, i.e. for each individual only SNPs overlapping with this set were taken, as this array is enriched for ancestry-informative non-selectable markers [24, 21]. We used the reference dataset from Elhaik et al. [21] as a base and added additional entries using supervised ADMIXTURE [25] analysis. Mean admixture coefficients were computed for each population in the database(see Elhaik et al. [21] for details). In the Dodecad Ancestry Project synthetic "zombies" are generated from the ADMIXTURE components. The concept of "reconstructed hypothetical ancient-like individuals" is similar to ancestral population used in our analysis. Here is the brief description of the approach:

1. Find allele frequencies of putative ancestral populations:

    • run ADMIXTURE [25] analysis in unsupervised mode on the entire reference dataset (possibly several times);

    • use CLUMPP [26] software to align and find consensus between *.P* matrices resulting from different runs and create a single *.P* matrix ($L \times K$, where $L$ is the number of loci, $K$ is the chosen number of putative ancestral populations).

2. For each $k = 1...K$, create ($m \approx 15$) individual genotypes by sampling the genotype at each locus $j = 1...L$ independently from binomial distribution ($n = 2$, $p = P(j, k)$ ). Genotype here is understood as number of copies of specific allele (0,1,or 2). These are the "zombie" genotypes, i.e. they represent a likely genotype of an individual from an ancestral population.

Following prior work of Elhaik et al. [24, 21], the resulting admixture coefficients were obtained from ADMIXTURE [25] analysis on an individual genome relative to $K = 9$ putative ancestral populations representing the genetic diversity of different geographic regions. This selection allows for direct comparison with prior work. However, larger values of $K$ are feasible to consider. We will continue inclusion of additional world-wide reference samples and experimenting with the number of components to achieve optimal performance of *reAdmix*.

33K Dataset

An additional reference dataset was constructed from microarray genotyping data on various worldwide populations. This dataset contains a smaller number of ancestry-informative markers, but a larger number of reference populations available in literature. This dataset is enriched for Native American, Chukotko-Kamchatkan, Siberian populations, as well as populations from South and North Caucasus. GenoChip ancestry-informative markers were selected in all datasets. Filtering of the resulting dataset was performed using the PLINK software [27] with the following criteria: maximum missing rate per SNP marker was 5%; maximum missing rate per individual was 50 (it was set so high to accommodate some important populations). The final dataset contained $1,564$ individuals from 86 populations and $33,039$ SNPs. We used unsupervised ADMIXTURE [25] analysis for $K$ ranging from 2 to 20. For each value of K, 100 admixture analysis runs were generated with different random seeds. The best run was chosen according to the highest value of log likelihood. We selected $K = 14$, since this number of components is high enough to provide the desired resolution, but at the same time is free of complicated ancestral populations substructure, that appears at higher values of $K$. Ten-fold cross-validation (CV) plots and admixture coefficients for various values of $K$ are shown in the Supplementary Materials.

## *reAdmix* approach

Instead of attempting to solve an "exact admixture" problem, we aim to find the smallest subset of populations whose combined admixture components are similar to those of the individual within a small tolerance margin. The reason for this is that the admixture proportions we use cannot be considered exact neither for the reference populations that consists of certain heterogeneity nor for the test individual, because the observed admixture proportions are merely maximum likelihood estimates, which may fail to accurately represent the actual proportions of ancestral genomes. Geometrically speaking, we seek to find a small subset of population points, such that their convex hull is adjacent to the test point in terms of maximum distance, defined as the maximum difference in the absolute values of two admixture coefficient vectors. The *reAdmix* algorithm solves this problem in two modes: conditional and unconditional. The conditional mode starts the search from one or more populations *a priori* provided by the user, whereas in the unconditional mode, no information is available.

The *reAdmix* algorithm consists of three phases (see Figure 2):

1. Iteratively build the first candidate solution, increasing the size by one population at each iteration, according to a criterion discussed below, until a maximum number of ancestral populations is found. The maximum number of the ancestral populations is a parameter which is defined using prior information about the ancestry composition, and roughly corresponds to the time-frame in question, represented by the number of generation. For example, to find the origin of one's grandparents the maximum number should be set to four, however the results may be like those of individual $T$ that in the simplest scenario may indicate common origins to two grandparents. Improve the candidate solution by exchanging populations in the solution for ones outside the solution space, if this substitution reduces the error.

2. Generate the predefined number $M$ of additional candidate solutions randomly and apply the Differential Evolution (DEEP) stochastic optimization technique to the combined set of the first and additional candidate solutions. The DEEP method is run for the $G_{max}$ predefined number of iterations using the objective function (3) described below that estimates the admixture proportions. The resulting set of $M+1$ solutions is subjected local optimization over all populations close to the obtained set. This resolves the problem of misplacing related populations such as Belorussian, Russian, and Ukrainian.

3. The populations that have stable membership in the solution across

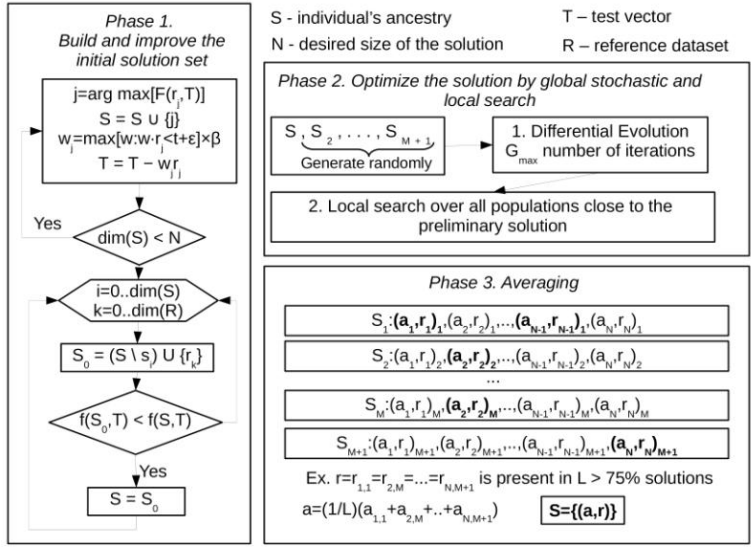

Figure 2: Flowchart of *reAdmix*.

the set, that is, are part of solution in at least 75% of cases, should be identified and reported, with their averaged estimates of admixture proportion.

## Notation

Let the reference dataset $R = (r_{ik})$ denote the matrix of admixture proportions of populations with respect to putative ancestral populations. We refer to the rows $r_i = (r_{i,1}, \ldots, r_{i,K})$ of matrix $R$ as population vectors. Let the admixture proportions of a test sample be denoted as $T = (t_1, \ldots, t_K)$. Let $S$ denote the solution vector, i.e. tuple of indices of populations that are present in test sample's admixture, and $A = (a_1, a_2, \ldots, a_p)$ the corresponding vector of mixture proportions to estimate. The $K$-component vector

$P = a_1 r_{s(1)} + a_2 r_{s(2)} + \cdots + a_p r_{s(p)}$ is the approximation of $T$.

## *reAdmix* algorithm description

<u>Initialization</u>. The set of populations present in individual's ancestry ($S$) is either empty (unconditional mode) or contains modern-day populations (conditional mode), provided by the user. Vector of proportions $A$ is undefined. Set $T_0 = T$, copy of the original test vector, as $T$ will change throughout the algorithm.

<u>Phase 1</u>. Build and improve the initial solution set.

1. Repeat the following steps until desired size of the solution set is reached:

    - Find the population vector with the highest *affinity score* (1) (see below) with respect to the current value of the test vector $T$, $j = \text{argmax}(\ F(r_j, T))$ .

    - Append this population to the solution set $S = S \cup \{j\}$.

    - Calculate the weight of the population vector to be proportional to the maximal possible (account for possible error) $w_j = \max[w : w \cdot r_j < t+\varepsilon] \times \beta$, where the scaling factor $\beta$ is empirically determined.

    - Subtract from the test vector $T$ the product of the population vector and its weight: $T = T - w_j r_j$ .

2. Improve the initial solution set by swapping populations with those outside of it. For all populations $x$ in the current solution and for all $y$ outside the solution, replace $x$ with $y$, if the change reduces the error.

<u>Phase 2</u>. Optimize the solution by global stochastic (1) and local search (2).

1. Stochastic step: The initial solution is combined with $M$ randomly generated vectors of populations' indices of the same size. Differential Evolution Entirely Parallel (DEEP) method is applied to this set of putative solutions for $G_{max}$ number of iterations. This makes it possible to identify the alternative combinations of populations that provide the lesser or the same error value as DEEP accepts only those substitutions in the parameter vectors that reduce the value of the objective function.

2. Local optimization: After obtaining the preliminary solution, a local optimization over all populations close to the preliminary solution is carried to identify the best possible solution. This step selects between related populations (e.g. Belorussian, Russian, and Ukrainian) that could have been misplaced in previous steps.

Phase 3. Averaging. To make a reliable estimate, the populations that have stable membership in the solutions across the set, that is, are part of solution in at least 75% of cases, should be identified and reported, with their averaged estimates of admixture proportion. We recommend to average across least $M = 10$ solutions to achieve stable results. The remaining populations should be considered potential contributing populations that may be grouped and reported as a regional population (e.g., South Europeans).

## Affinity score

Affinity score of a vector $P$ to a test vector $T$

$$F(P, T) = \arg\min_{\alpha} L(d(\alpha)) \qquad (1)$$

is the value of the weight $\alpha$ such that the difference between prediction and true value of test vector $d = T - \alpha P$ minimizes the loss function

$$L(d) = \sum_{i=1}^{K} d^2_i + \sum_{i: d_i < \varepsilon} (1 + 2d_i) \qquad (2)$$

The goal of the second term is to penalize for inclusion of too many ancestral populations (i.e. when $\alpha P_i > T_i$).

## Objective function

Population weights are considered optimal if they minimize the absolute error of the solution, i.e. the maximum absolute error between the approximation defined by $S$, $A$, and $T$. The function finds proportions $A = (a_1, a_2, \ldots, a_p)$ corresponding to the elements of approximation defined by $S = (s_1, s_2, \ldots, s_p)$ such that the absolute solution error

$$f(S, T) = \min_{A=(a_1, a_2, \ldots, a_P)} \max_{k=1 \ldots K} P - T, \qquad (3)$$

where $P = a_1 r_{s(1)} + a_2 r_{s(2)} + \cdots + a_p r_{s(p)}$, is minimal. The minimization of absolute error is an instance of Chebyshev approximation linear programming problem. To solve it we use *lpSolve* package [28].

## Differential Evolution Entirely Parallel Method

Recently, many promising optimization techniques have been developed based on the Differential Evolution originally proposed by Storn and Price in [29, 30]. To solve our optimization problem, we adopted the Differential Evolution Entirely Parallel (DEEP) method [31] incorporating into the original algorithm such enhancements found in the literature as the possibility to take into account a value of the objective function for each parameter vector at the recombination step [32], and to control the diversity parameter vectors by the adaptation of the internal parameters [33]. DEEP starts from a set of the randomly generated parameter vectors $q_i$, $i = 1, ..., NP$. The size of the set $NP$ is fixed. The first trial vector is calculated by:

$$v = q_{r1} + S(q_{r2} - q_{r3})$$

where $q.$ is the member of the current generation $g$, $S$ is a predefined scaling constant and $r_1$, $r_2$, $r_3$ are different random indices of the members of population. The second trial vector is calculated using "trigonometric mutation rule" [32].

$$\begin{aligned}z &= q^{r_1} + \frac{q_{r2} + q_{r3}}{3} + (\varphi_2 - \varphi_1)(q_{r1} - q_{r2}) \\ &+ (\varphi_3 - \varphi_2)(q_{r2} - q_{r3}) + (\varphi_1 - \varphi_3)(q_{r3} - q_{r1})\end{aligned}$$

where $\varphi_i = |F(q_{r_i})|/\varphi^*$, $i = 1, 2, 3$, $\varphi^* = |F(q_{r1})| + |F(q_{r2})| + |F(q_{r3})|$, and $F(x)$ is the main objective function to be minimized. The combined trial vector in case of binomial recombination type is defined as follows:

$$w_j = v_j * (U_j(0,1) < p) + z_j * (U_j(0,1) < 1 - p)$$

where $U_j(0, 1)$ is a random number uniformly distributed between 0 and 1 and $p$ is the probability of crossover. In case of the exponential type of recombination the first trial vector $v$ is used continuously while random number is less than $p$.

Several different objective functions can be used to decide if the trial vector will replace the current one in the set. The trial vector is accepted if the value of the main objective function decreased. In the opposite case the additional objective functions are considered if they are defined. The trial vector replaces the current one if the value of any other objective function is better, and a randomly selected value is less than the predefined parameter for this function.

It is worth noting that the DEEP method was previously successfully applied to several systems biology problems [34, 35, 36]. The distinctive features of the DEEP method are the flexible selection rule for handling multiple objective functions and substitution strategy that takes into account the number of iterations between updates of each parameter vector. Several oldest vectors are substituted with the same number of the best ones after predefined number of iterations. Different types of experimental observations or a priori knowledge can be included in one fitting procedure using the new selection rule. We are currently developing a nonparametric [37, 38] version of the *reAdmix* approach.

The algorithm was implemented in C programming language as the software package with interface that allows a user to formulate the objective function using different computer languages widely used in biomedical applications, such as Octave, R, etc. The control parameters of the algorithm are defined in the data file that uses the INI-format. The package provides the simple command line user interface.

One of the parameters of the algorithm determines the number of parallel threads used to calculate the objective function. We utilized the Thread Pool API from GLIB project https://developer.gnome.org/glib/ and constructed the pool with the defined number of worker threads. The calculation of objective function for each trial vector is pushed to the asynchronous queue. The calculation starts as soon as there is an available thread. The thread synchronization condition is determined by the fact that objective function is to be calculated once for each individual in the population and on each iteration.

# Competing interests

The authors declare that they have no competing interests.

# Author's contributions

TT formulated the problem and conceptualized the method, DC and KK implemented the method, MH and MT developed the interface, PF and PT prepared the testing and reference datasets, TT, KK, and DC interpreted the results and wrote the paper.

# Acknowledgements

Funding: TT was supported by grants from The National Institute for General Medical Studies (GM068968), and the Eunice Kennedy Shriver National Institute of Child Health and Human Development (HD070996). KK was supported by the "5-100-2020" Program of the Ministry of Education and Science of the Russian Federation.